\documentclass[12pt,preprint]{aastex} 

\shorttitle{ Loop Width Comparison }
\shortauthors{Chastain and Schmelz}

\begin{document}

\title{A Comparison of EIT and TRACE Loop Widths}

\author{
S.I. Chastain\altaffilmark{1}
J.T.\ Schmelz\altaffilmark{1,2,3}
} 
\altaffiltext{1}{Physics Department, University of Memphis, Memphis, TN 38152}
\altaffiltext{2}{USRA, 7178 Columbia Gateway Drive, Columbia, MD 21046, jschmelz@usra.edu}
\altaffiltext{3}{Arecibo Observatory, HC-3 Box 53995, Arecibo PR 00612}



\begin{abstract}
In this study we have compared coronal loops in the Extreme ultraviolet Imaging Telescope (EIT) on-board the Solar and Heliospheric Observatory (SOHO) with coronal loops from the Transition Region and Coronal Explorer (TRACE). The purpose of which is to quantitatively and qualitatively examine the effects of spatial resolution on the width of coronal loops and implications for how a coronal loop is defined. Out of twenty-two loop sections analyzed, we find that none of them were resolved in EIT and none of them were close to the width of the TRACE loops. These findings suggest that coronal loops are unresolved in EIT. We also find examples of how unresolved loops can be quite misleading. We have also found that many of the TRACE loops that we have analyzed may be unresolved as well. Our findings emphasize the importance of studying loop width in order to better understand coronal loops and also emphasize the need for instruments with higher spatial resolution. 
\end{abstract}


\section{Introduction}

The launch of Skylab in 1973 allowed for space-based observations of the corona in the soft x-ray and extreme ultraviolet wavelengths. The observations from the extreme ultraviolet and soft x-ray telescopes on Skylab showed loops filling the corona, and helped solar physicists create models for the corona that account for the wide variety of features visible. Subsequent early models such as Rosner et al. (1978) emphasize the role of coronal loops in heating the corona, and assume that each loop is a single magnetic flux tube with plasma in hydrostatic equilibrium. Beginning in the 1990s with the launches of the Soft X-ray Telescope on Yohkoh, EIT on SOHO, and TRACE, many loops were analyzed using these assumptions along with filter ratio analysis (e.g. Aschwanden et al. 2000). These results seem to confirm the early assumptions made about loops. However, in the early 2000s some observations began to conflict with previous results and assumptions.  Winebarger et al. (2003a,b) found electron densities in loops that exceed that which was predicted by models of loops with plasma in hydrostatic equilibrium and uniform heating. Furthermore, Warren et al. 2003 found loops that have a longer lifetime than that which would be expected from uniform heating. In addition, Schmelz et al. (2016) shows numerous examples of multithermal loops alongside a few isothermal loops as well. These results challenge the original assumptions about coronal loops. 

The study of the width of loops has a slightly shorter history. Klimchuk et al. (1992) was one of the first studies to quantitatively analyze loop width. They analyzed loops using the Soft X-ray Telescope on Yohkoh. This study along with Klimchuk (2000) found that, contrary to expectations, loops do not significantly vary in width along the length of the loop. These results were confirmed in the measurements taken from TRACE loops (Watko \& Klimchuk 2000). L{\'o}pez Fuentez et al. (2006) directly compare TRACE measurements with models made from measurements of magnetic flux using MDI on SOHO. Their findings reinforce the conflict between loop width and the models of the magnetic field. DeForest (2007) argues that TRACE loops may be composed of unresolved threads of around 10-100 km in size at hydrostatic equilibrium. With the launch of the High Resolution Coronal Imager in July of 2012, loops could be observed on length scales of nearly one-tenth of an arcsecond. Peter et al. (2013) analyze the results of the Hi-C images and compare them to AIA images in the same active region. They found that the AIA loops did not show significant substructure and that if loops are composed of unresolved threads, these threads would need to be less than 15 km in diameter. 

In light of these discoveries, it is reasonable to conclude that the original definition of a loop and the assumptions made along with it are no longer sufficient to describe the observations and theories involving multithreaded loops. One way of updating the definition of a loop is to say that a loop is a distinct configuration in an observation, and each loop is made up of strands that are elementary flux tubes. Although this definition holds up to the observations that have challenged previous definitions, it suggests that a loop in one instrument is not a loop in another instrument. This conterintuitive situation suggests an inconsistency of loops between observations and deserves some attention. How can it be that a loop in one observation is not a loop in another observation?

In order to fully explore this issue, we analyze loops both qualitatively and quantitatively by comparing loops in EIT and TRACE. If a loop was clearly visible in only one instrument, we describe what is visible instead of the loop. Every visible loop can be analyzed by plotting the loop width along the length of a segment of the loop (see figure 5, for example). For loops that appear in both instruments, we are able to put both instruments on the same plot. Comparing loops from EIT and TRACE in this manner allows us to answer broad questions, such as questions about the presence of a corresponding loop to a known loop, while also allowing us to answer specific questions about whether or not loops are resolved and the scale of the loop in question. 

\section{Observations}

The Transition Region and Coronal Explorer was launched on 2 April 1998 and completed its mission in June of 2010. TRACE was a part of the NASA Small Explorer mission. It was tasked with imaging the solar corona and transition region at a high angular and temporal resolution. TRACE features a spatial resolution of 0.5 arcseconds per pixel.  In addition to the many ultraviolet filters on TRACE, it also contains three filters focused on ionized iron: 171-\AA\ focused on Fe IX, 195-\AA\ focused on Fe XII, and 284-\AA\ focused on Fe XV. Because loops are both more visible and appear more distinctly in the 171-\AA\ filter, observations were limited to this filter.

In addition to TRACE, data were also obtained from the Extreme Ultraviolet Imaging Telescope. EIT was launched on board the Solar and Heliospheric Observatory on 2 December 1995 and completed its mission in July of 2010. SOHO was part of a collaboration between ESA and NASA to study the Sun from the deep core to the outer corona and solar wind.  SOHO circles the first Lagrangian point between Earth and the Sun for an uninterrupted solar view and constant communication with the Earth. EIT features a spatial resolution of 2.6 arcseconds per pixel. EIT contains four filters: 171-Å focused on Fe IX, 195-\AA\ focused on Fe XII, 284-\AA\ focused on Fe XV, and 304-\AA\ focused on He II. In addition to the reasons listed above, observations were limited to the 171-\AA\ filter in order to compare loops in the contrasting spatial resolutions of TRACE and EIT. 

TRACE data were obtained from the TRACE data center and the Virtual Solar Observatory website. Raw TRACE data were read, processed, and calibrated using the read\_trace and trace\_prep programs available in SolarSoft. Level Zero EIT data were obtained from the Virtual Solar Observatory website and the EIT data catalog.  EIT data were read, processed, and calibrated using the read\_eit and eit\_prep programs available in SolarSoft. 

Active regions of interest were selected based off of the presence of a loop against a clean background and the availability of data from both EIT and TRACE taken within 10 minutes of each other. 10 Loops were selected from EIT and matched up with the corresponding TRACE image, and 12 loops were selected from TRACE and matched up with the corresponding EIT image. Figures 1 and 2 show the active regions used and the selected loops. Table 1 lists the active regions used, the location of the regions, and the instrument from which the loops were first chosen. Loops were chosen from both on the limb and on the disk.

\section{Analysis}

Only loops with clean background were selected for loop width analysis. In order to analyze loop width, the same program and procedure as was used in Klimchuk (2000) was followed: loop sections were straightened and loop pixels were selected. A polynomial background fit was performed. In order to estimate error in loop pixel selection, a second analysis was performed for each loop with one additional pixel on either side of the loop. The end result is a standard deviation of the intensity profile, plotted as a function of position along the length of the loop, as shown in figure 5. If the loop cross section is circular and radiates uniformly, then the diameter is four times the standard deviation (Klimchuk 2015). For the images that were matched up with the selected loop, if a distinguishable loop was also present in the corresponding image, the width of the loop in the corresponding image was analyzed and compared with the width of the loop in the original image. If a distinguishable loop was not present in the corresponding image, then the image is described below.

There are several reasons why there may be no analyzable loop in the corresponding images. In six of the ten loops selected from EIT first, what appears to be one single loop in EIT is actually multiple features in TRACE. Problems related to features merging in EIT made up the largest category of issues related to the inability to analyze loop width. All of the loops selected from EIT first that could not be analyzed fell into this category. Among loops that were selected from TRACE first, there were two main issues that appeared in combination resulting in a lack of an analyzable loop. These issues can be described as `high background' and `not well defined.' Corresponding images with high background have bright material near the loop location that obscures the loop. Corresponding loops that are not well defined show an increase in brightness in the general area of the loop, but no clearly visible loop is present.

A total of twelve loops were selected from TRACE first. Of these twelve loops, seven could not be matched up with a single, clear loop in EIT. Five of the seven loops have visibility and background issues. The other two loops merge to make a single loop in EIT. Region C Loops 1 and 2 were selected from the same active region in TRACE. These loops merge into one loop in EIT. These loops bring to mind figure 2 from DeForest (2007). As seen in figure 6, the loop that appears in EIT is much narrower than EIT can resolve. This phenomenon shows the limitations of analyzing loops that are relatively narrow compared to the spatial resolution of the instrument. For Region C Loop 3, there is no loop visible in EIT, likely due to high background. In Region D Loop 2, the loop has high background. For Region D Loop 4, the loop is not well defined. For Region F Loop 2, the loop is too small and has high background.  For Region F Loop 3, the loop is not well defined and has high background.

Of the twenty-two loops analyzed, eight were able to be matched up with a corresponding loop. Even our best examples were less than two pixels in width. Nevertheless, some clear trends can be seen in comparing the three loops with the worst correspondence between instruments with the three loops with the best correspondence between instruments (see figure 5). Loops that are less than or around one pixel in width in the instrument observed can be considered unresolved. Our best examples were clearly resolved in TRACE with Region A Loop 2 being resolved in both TRACE and EIT. In contrast, none of our worst examples appear to be well resolved by TRACE or EIT. However, even though Region A Loop 2 is resolved in EIT, the loop width in EIT more than four times that of the TRACE width of the same loop and is not much larger than a single EIT pixel. In fact, no loop had similar width between instruments. The EIT loops we have analyzed all appear to be unresolved. This finding suggests that the EIT loop seen in Region A Loop 2 is not a good comparison to the loop by the same name in the TRACE image. This finding may have implications for the spatial resolution required to resolve most coronal loops.

\newcommand{\ltapprox}{\stackrel{<}{\sim}}

\section{Discussion}

Of all the twenty-two loops selected for analysis, thirteen loops could not be compared to an equivalent loop in the other instrument (EIT or TRACE). A majority of the corresponding images show features merging or disappearing. These kinds of loops are the sorts of loops that cause problems with not only analysis, but also defining loops as well. For example, in Region C Loops 1 and 2, there are two loops present in TRACE and only one loop present in EIT. At this point the observer must decide if they are looking at a single, different loop in the EIT image or if it is conceptually the same two loops. An important factor in this issue is whether the loops are resolved or unresolved in the observation. In order to accurately communicate information on coronal loops, it is important that loops are distinguished by whether they are resolved or unresolved. Distinguishing between unresolved and resolved loops may determine the implications of an analysis of loops.

Issues in the analysis of coronal loops in EIT have been previously noted (see, e.g., Schmelz et al. 2003). Therefore, it should not come as a surprise that there were issues with analyzing the width of coronal loops in EIT. To be more specific, of the loops that were able to be compared quantitatively, none of the EIT loops were much larger than the size of a single pixel in EIT. This finding suggests that EIT loops are unresolved. Of the loops analyzed, there was one example in particular that appeared to be resolved in a qualitative analysis, but appeared to be unresolved after a quantitative loop width analysis. Region A Loop 2 appears clearly in EIT and can easily be matched up with its corresponding loop that appears in TRACE. However, upon analyzing the width of this loop in EIT, it is clear that this loop is not much wider than a single pixel in EIT. The width in TRACE is, in contrast, much less than this width, near two TRACE pixels. 

We return to Region C Loops 1 and 2, loops which demonstrate the case of multiple loops in one instrument merging to form a single loop in another instrument. DeForest (2007) presented this situation in their figure 2. They were pointing out that TRACE loops could be unresolved. In our case, it is instead a loop an EIT that is clearly two loops in TRACE. Worth noting is that the analysis of the combined loop in EIT (see figure 6) shows a resulting loop that is less than the width of a single EIT pixel. The claim that loops are unresolved appears to hold for EIT loops. In TRACE, however, at least some coronal loops appear to be fairly well resolved. In our analysis, we have seen loops that were larger than the resolution limit of TRACE. Also in their comparison between AIA and Hi-C, Klimchuk (2015) analyzes four loops that are present in the 193-\AA\ filter. They find that width measurements are quite similar between the two instruments. These four AIA loops are fairly well resolved. Since the spatial resolution of TRACE is slightly better than that of AIA, then at least some TRACE loops ought to be resolved. It is worth noting that many TRACE loops were near the size of a pixel. This may mean that, like EIT, there are some loops that are unresolved in TRACE. However, this is also consistent with Peter et al. (2013) who observed some loops that were well below the resolution of AIA and would also be below the resolution of TRACE. 

The implications for analysis of loops are quite clear for EIT: EIT loops are not suitable for quantitative analysis. As discussed previously, this is already well known. The findings for AIA and TRACE are a bit less clear. Our findings along with the findings of Klimchuk (2015) suggest that analyzing loops in TRACE ought to provide good results in most cases, since many loops ought to be resolved. However, it should not be surprising if in an analysis there are some loops that do not provide good results. Some loops will be below the ability of AIA or TRACE to resolve. These kinds of results are consistent with the analysis of Schmelz et al. (2016). They analyzed many loops with good results in a very large analysis, but also found one loop that did not produce good results in analysis. Their results are supported by our finding that many loops are resolved or nearly resolved in TRACE, TRACE being close to AIA in spatial resolution, but that a few loops may be unresolved.

If EIT loops are, in general, unresolved and some TRACE loops are resolved and others are not, supports the assertion of Klimchuk, 2015 that loops have a preferred spatial scale. This assertion is reinforced by the fact that none of the loop sections analyzed in TRACE were very much larger than two pixels in width. Our findings show an upper limit to the width of coronal loops and that such a limit may be near two pixels in TRACE or about 3000 km. This upper limit is smaller than a single EIT pixel, suggesting that coronal loops ought to be unresolved EIT. In light of this finding, it is perhaps unwise to refer to coronal loops in EIT without specifically mentioning that they are ``unresolved loops." 

In our statement of the definition of a coronal loop, we specifically noted that a loop is a distinct configuration in an observation. We have shown several examples of issues that arise from unresolved loops. While this definition of a loop is not inaccurate, not giving any information on whether observed features are resolved or not may lead to misleading results. For this reason, in instruments such as EIT where loops are visible but it is fairly clear that loops are unresolved, these loops should be noted as unresolved. These unresolved loops may be quite different from loops visible in TRACE. While we are not sure exactly how many loops are resolved in TRACE, or AIA for that matter, many loops may be resolved or nearly resolved loops and until another instrument with better spatial resolution exists to verify, these loops are simply loops.

\section{Conclusions}

Our study begin with an interest in the nature of loops in varying spatial resolutions. In our inquiry we posed the following question: How can it be that a loop in one observation is not a loop in another observation? In order to find an answer to this question we examined and compared loop sections both quantitatively and qualitatively. We found that even in loops that appear to be the same in EIT and TRACE, the width of the loops were quite different. We also found that none of the loop sections analyzed in EIT were much wider than one pixel in EIT. We found a few examples of TRACE loops that were clearly larger than a TRACE pixel. However, we also found several that were not, and could potentially be unresolved or poorly resolved. 

With these results we can now answer our original question. A loop can be a loop in one observation and not in another observation when it is unresolved. While seemingly a simple conclusion, it is a quite important one. We have seen examples in our results of how misleading a loop can appear to be when it is unresolved. In addition, by examining the width of coronal loops we can get an idea of whether or not a loop is unresolved without necessarily needing another instrument with better spatial resolution. These findings highlight the importance of loop width studies when analyzing coronal loop widths.

The authors would like to thank Jim Klimchuk for sharing software and useful discussions. This work was part of a senior capstone project at the University of Memphis. It was inspired by discussions at the Coronal Loops Workshops. 

{}

\clearpage

\begin{deluxetable}{llllc}
\tabletypesize{\scriptsize}
\tablewidth{0pt}
\renewcommand{\tabcolsep}{5.0pt}
\tablecaption{Active Regions}
\tablehead{
\colhead{ } & \colhead{Region} & \colhead{Date} & \colhead{Coordinates} & \colhead{Loops Selected From} 
}
\startdata
A	&	8759	&	1999 November 06	&	N10 E64	&	EIT	\\
B	&	8820	&	2000 January 13	&	S14 W22	&	TRACE\\
C	&	8910	& 	2000 March 24	&	N13 W89	&	TRACE\\
D	&	9033   &	2000 June 11	&	N22 E04 	&	TRACE	\\
E	&	9077   &	2000 July 11	&	N17 E30	&	EIT	\\
F	&	9574   &	2001 August 12	&	S04 W24	&	TRACE
\enddata
\end{deluxetable}
 
\clearpage

\begin{figure}
\epsscale{.9}
\plotone{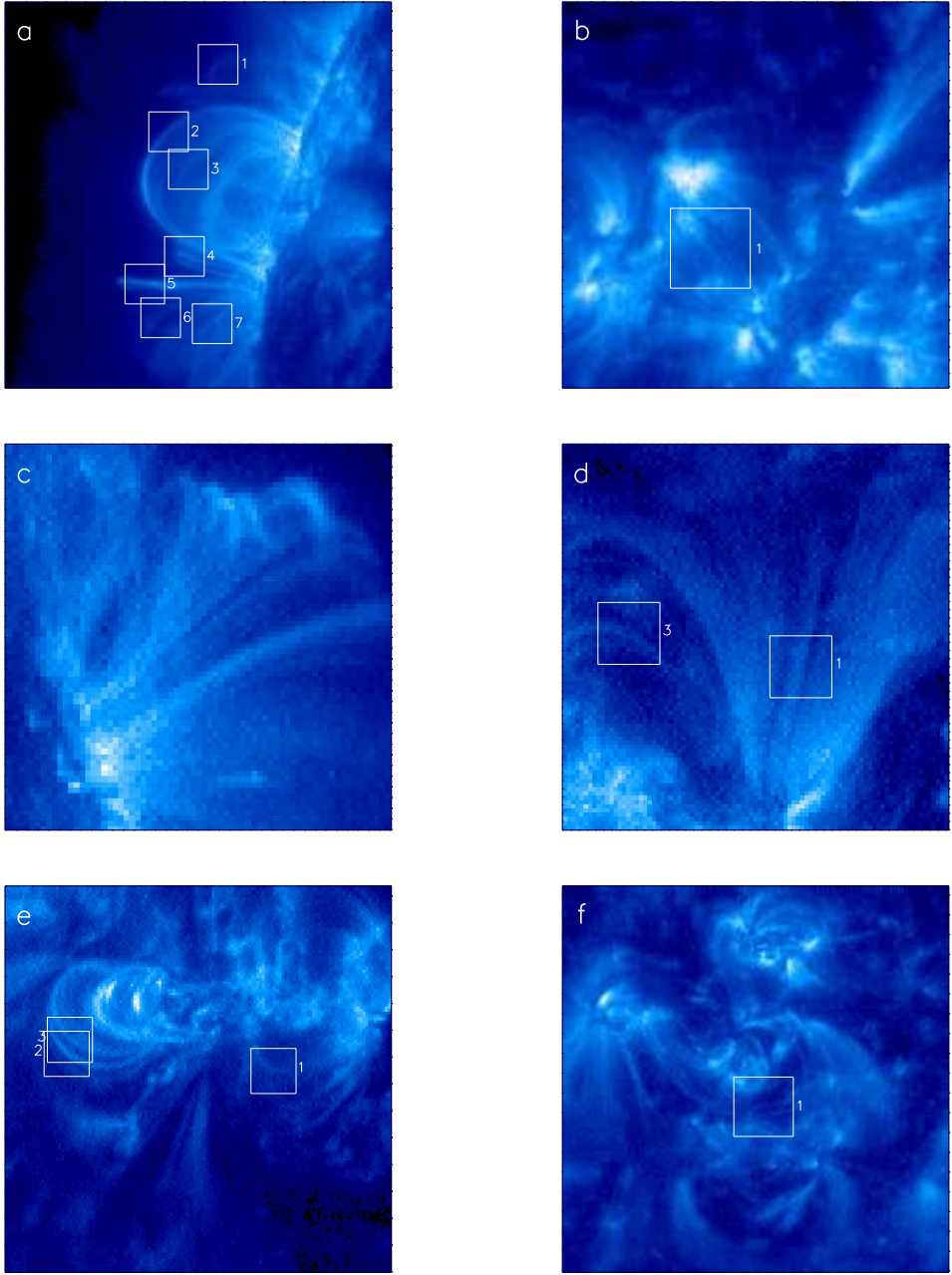}
\caption{EIT 171-\AA\ filter images of the active regions described in Table 1. Each loop segment that has been analyzed is designated with a small white box and a number.
}
\end{figure}

\clearpage
\begin{figure}
\epsscale{.9}
\plotone{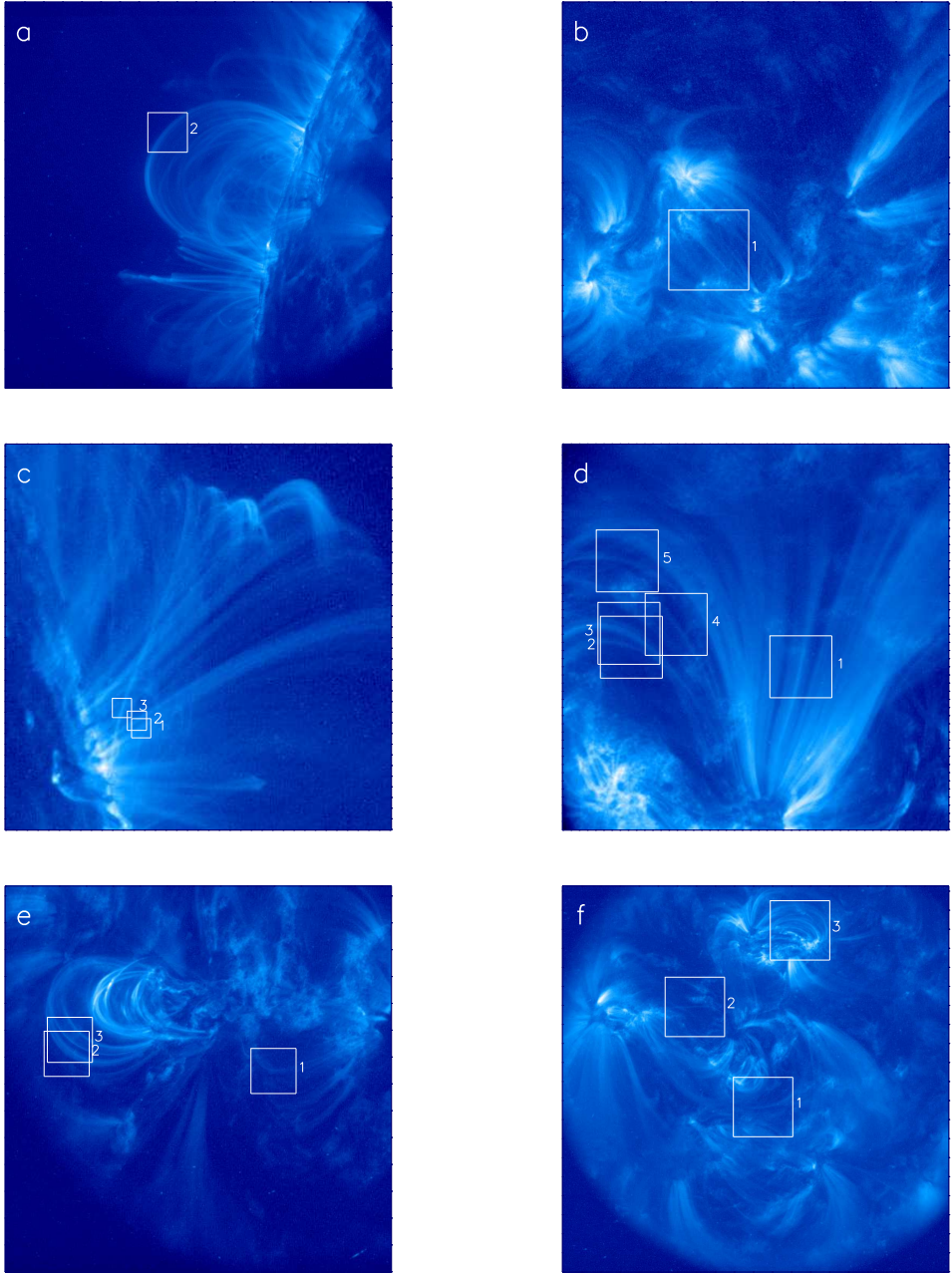}
\caption{TRACE 171-\AA\ filter images with a format similar to Figure 1.
}
\end{figure}

\clearpage
\begin{figure}
\epsscale{.9}
\plotone{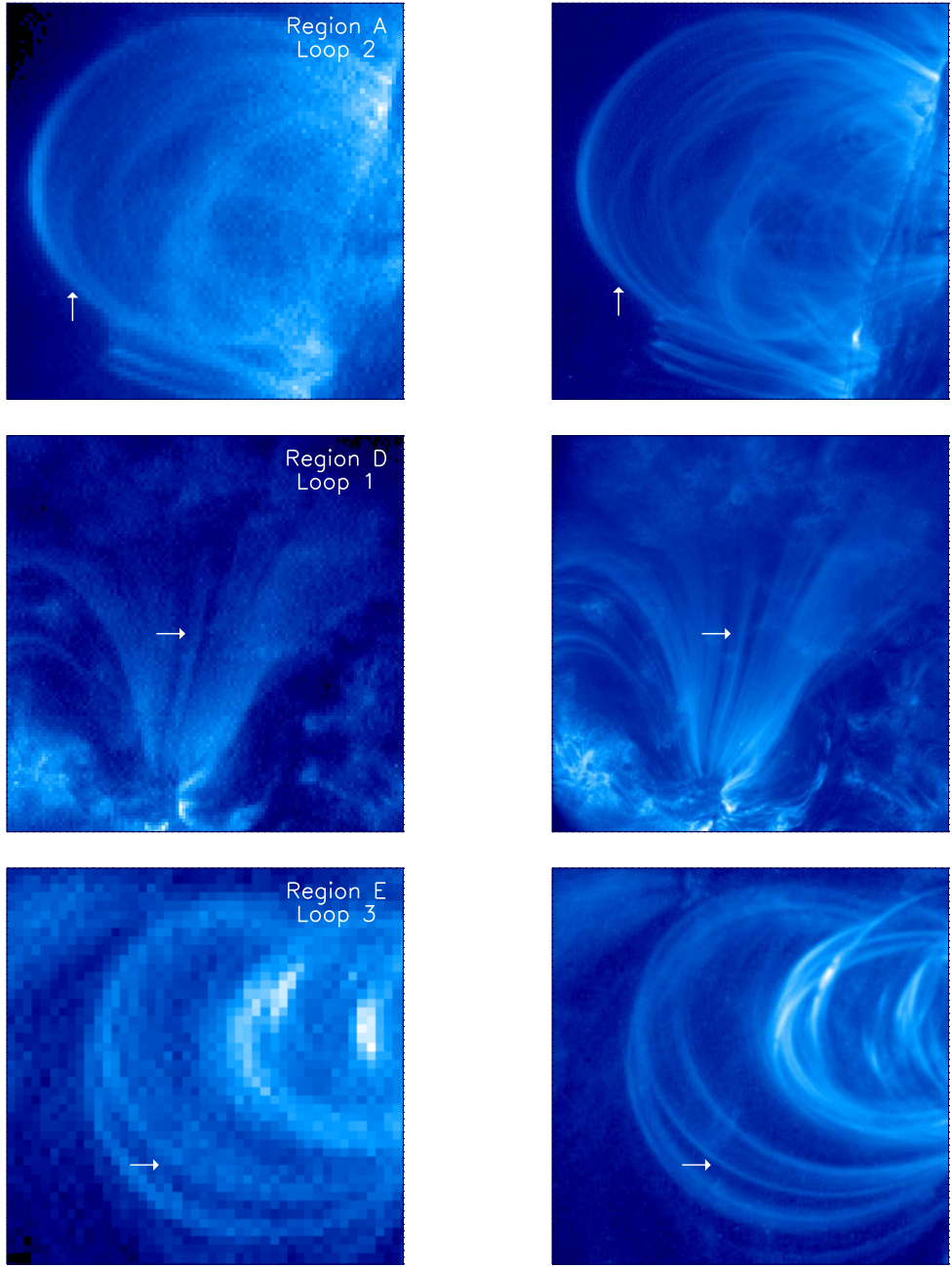}
\caption{Close-up images showing the best correspondence between EIT and TRACE loops. Each loop is indicated by a white arrow. EIT images are on the left, and TRACE images are on the right.
}
\end{figure}

\clearpage
\begin{figure}
\epsscale{.9}
\plotone{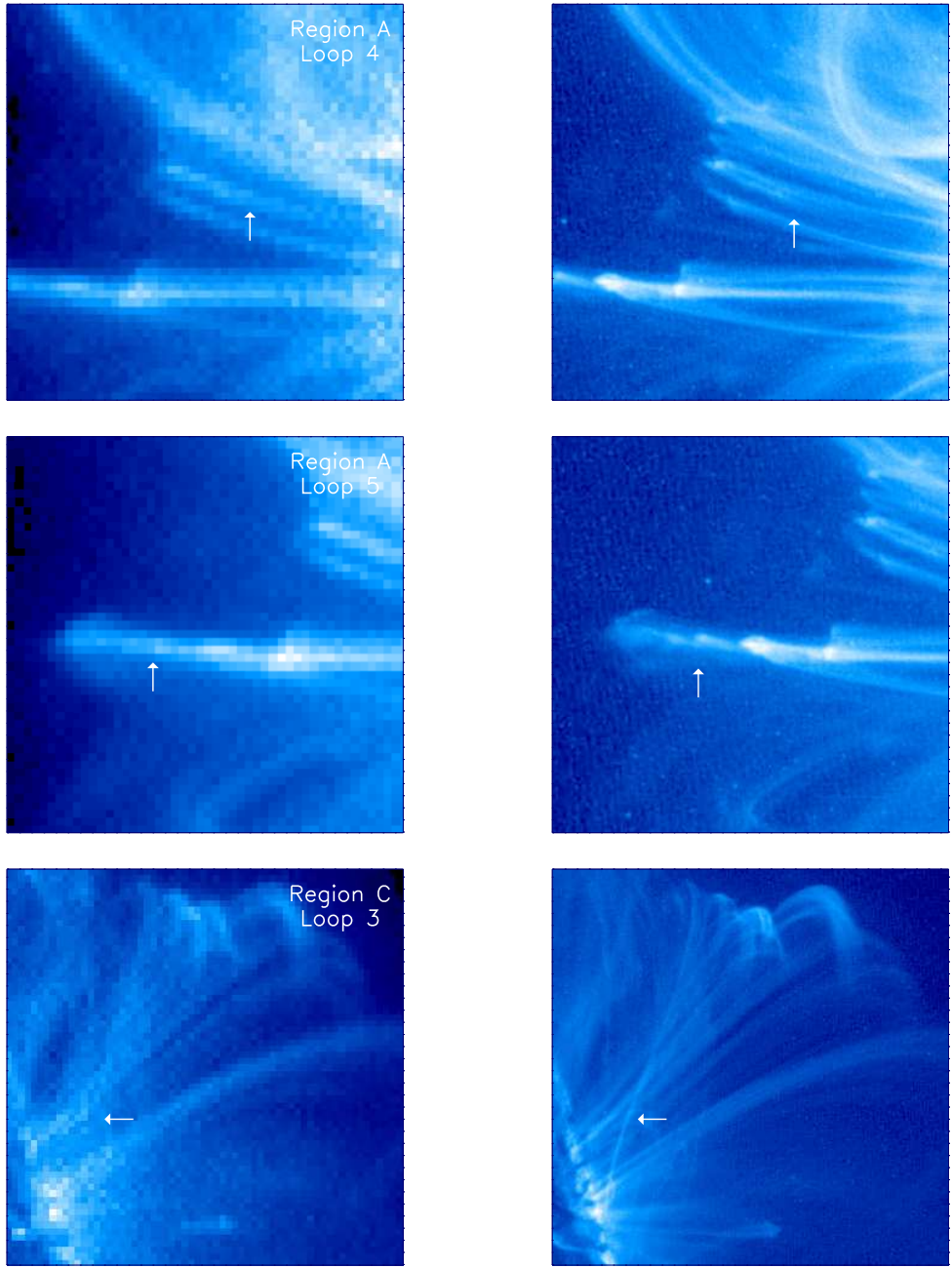}
\caption{Close-up images with a similar format to Figure 3 showing the worst correspondence between EIT and TRACE loops. 
}
\end{figure}

\clearpage
\begin{figure}
\epsscale{.9}
\plotone{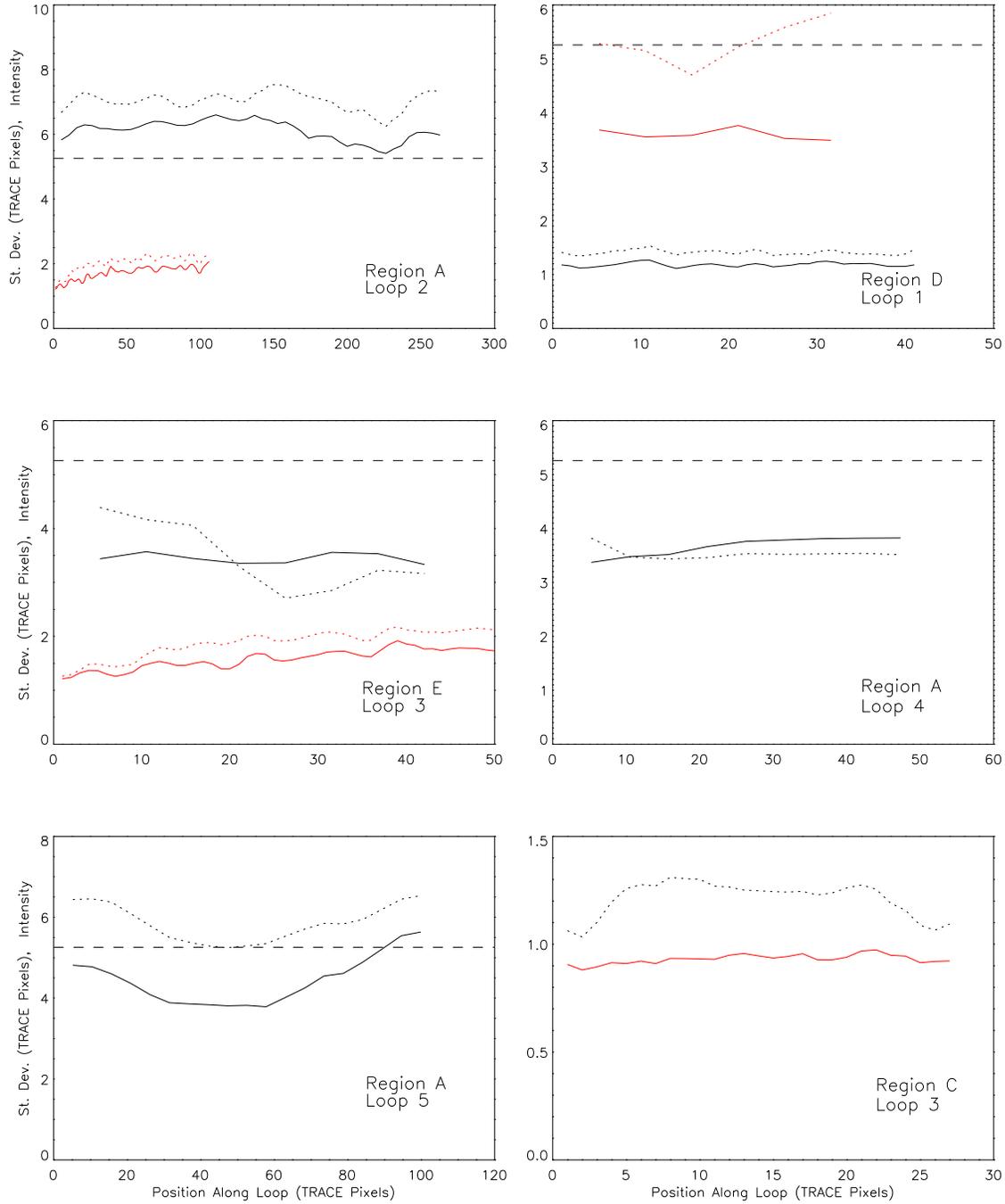}
\vspace{20pt}%
\caption{ 
For each panel, the each line shows the standard deviation of the intensity profile along the length of the observed loop segment. The black line is from loops observed in EIT. The red line is from loops observed in TRACE. Data from EIT has been converted to equivalent TRACE pixels. The dashed line represents a single EIT pixel. The dotted lines represent the error pass.}
\end{figure}

\clearpage
\begin{figure}
\epsscale{.9}
\plotone{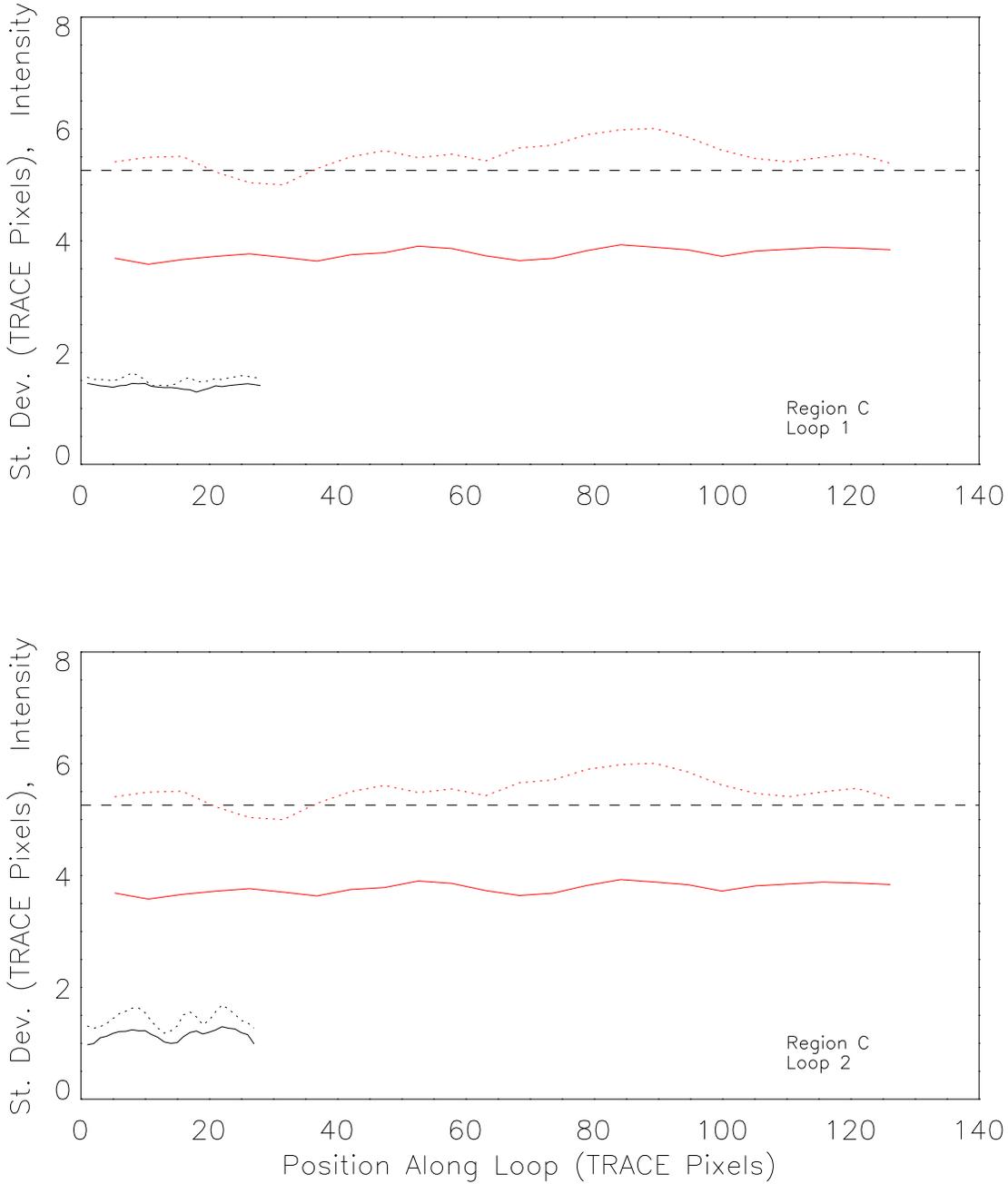}
\vspace{20pt}%
\caption{ A plot similar to Figure 5 for Region A Loops 1 and 2. The two TRACE loops merge to form a single EIT loop. The EIT data for the merged loop is plotted in black. TRACE data is plotted in red. 
}
\end{figure}

\end{document}